\documentclass[11pt,a4paper]{article}
\usepackage{amsmath}
\usepackage{enumerate}
\usepackage{amssymb}
\usepackage{graphicx}
\usepackage{multirow}
\usepackage{xcolor}
\usepackage{tablefootnote}
\usepackage{threeparttable}
\usepackage{bm}
\usepackage[normalem]{ulem}
\usepackage{jcappub}

\begin{document}

\title{Cosmic Chronometers with Photometry: a new path to $H(z)$}
\author[a,b]{Raul Jimenez,}
\author[c,d]{Michele Moresco,}
\author[a,b]{Licia Verde,}
\author[e,f,g]{Benjamin D.~Wandelt}

\affiliation[a]{ICC, University of Barcelona, Mart\' i i Franqu\` es, 1, E08028
Barcelona, Spain}
\affiliation[b]{ICREA, Pg. Lluis Companys 23, Barcelona, 08010, Spain.}
\affiliation[c]{Dipartimento di Fisica e Astronomia ``Augusto Righi'', Universit\`a di Bologna, Viale Berti Pichat 6/2, I-40127, Bologna, Italy}
\affiliation[d]{INAF - Osservatorio di Astrofisica e Scienza dello Spazio di Bologna, via Gobetti 93/3, 40129 Bologna, Italy}
\affiliation[e]{Sorbonne Universit\'e, CNRS, UMR 7095, Institut d'Astrophysique de Paris, 98 bis bd Arago, 75014 Paris, France.}
\affiliation[f]{Sorbonne Universit\'e, Institut Lagrange de Paris (ILP), 98 bis bd Arago, 75014 Paris, France.}
\affiliation[g]{Center for Computational Astrophysics, Flatiron Institute, 162 5th Avenue, 10010, New York, NY, USA.}

\emailAdd{raul.jimenez@icc.ub.edu}
\emailAdd{michele.moresco@unibo.it}
\emailAdd{liciaverde@icc.ub.edu}
\emailAdd{bwandelt@iap.fr}

\abstract{
We present a proof-of-principle determination of  the Hubble parameter $H(z)$ from photometric data, obtaining a determination at an  effective redshift of $z=0.75$ ($0.65<z<0.85$) of $H(0.75) =105.0\pm 7.9(stat)\pm 7.3(sys)$ km s$^{-1}$ Mpc$^{-1}$, with 7.5\% statistical and 7\% systematic (10\% with statistical and  systematics combined in quadrature) accuracy. This is obtained in a  cosmology model-independent fashion, but assuming a linear age-redshift relation in the relevant redshift range, as such, it can be used to constrain arbitrary cosmologies as long as $H(z)$ can be considered slowly varying over redshift.
In particular, we have applied a neural network, trained on a well-studied spectroscopic sample of  140 objects, to  the {\tt COSMOS2015} survey  to construct a set of 19 thousand near-passively evolving galaxies and build an age-redshift relation. The Hubble parameter is given by the derivative of the red envelope of the  age-redshift relation. This is the first time the Hubble parameter is determined from photometry at $\lesssim 10$\% accuracy.  Accurate $H(z)$ determinations could help shed light on the Hubble tension;   this study shows that photometry, with a reduction of only a factor of two in the uncertainty, could provide  a new  perspective on the tension.}
 
\maketitle

\section{Introduction}
\label{sec:intro}

The spectacular success of the standard model of cosmology, the so-called Lambda Cold Dark Matter ($\Lambda$CDM) model, relies on observations of the early Universe  via the  cosmic microwave background, standard rulers (calibrated on early-time physics), and standard candles (Supernovae type Ia).  The precision tests of this model, enabled by the avalanche of  cosmological data of the past two decades,  instead of answering the open questions about the nature of the accelerated expansion  and the  constituents of the Universe, have surfaced some discrepancies or ``tensions'' \cite{tensions}. Whether these tensions are an indication of a possible limitation of the standard cosmological model or are the symptom of  low-level unknown systematic errors is still a matter of much debate in the community. 
It is well-known that each probe has its own strengths and weaknesses, is sensitive to specific physical processes, and is affected by a specific set of systematics: it is the combination of multiple, complementary,  probes that contributes to ensure the robustness of any finding as  clearly highlighted by \cite{Albrecht2006}. The effort of the scientific community has since proceeded in this direction; the diversity of different methods and probes is increasingly gaining relevance and importance, and looking for new independent cosmological probes may be key to address the $\Lambda$CDM ``tensions''. 

A recent review~\cite{Morescoreview} has presented the advances in developing alternative, emerging,  probes that can be used to test key aspects of the cosmological model and constrain (or even possibly detect) deviations from it.

One of these probes is the cosmic chronometer (CC) method~\cite{JimenezLoeb} which can be used to measure the expansion rate of the Universe $H(z)$. To this date,  this is the only truly cosmology-independent method to measure $H(z)$ without relying on assuming the cosmological model. Cosmic chronometers refers to a population of  old  passively evolving objects (galaxies) all with synchronous stellar populations: cosmic chronometers started forming their stars all approximately at the same time and much earlier than their time of observation.  The differential ages of such population observed across redshifts yield a determination of $H(z)$.  

Until now, the CC method has been applied to high signal-to-noise spectroscopic surveys~\cite{Simon,Stern,Moresco1,Moresco2,Borghi2,Jiao,Tomasetti} of passively evolving galaxies. The reason is  that selecting the CC (objects that populate the  red envelope, or the edge of the age-redshift relation) and  measuring differential ages robustly, precisely and accurately, is a very  challenging task due to the effects of other (stellar) parameters that also determine the integrated light of the stellar population of a galaxy. By focusing on particular absorption features of the rest-frame optical spectrum~\cite{Moresco1}, it has become possible to  estimate $H(z)$ in the redshift range $0.05 < z < 2$ (see~\cite{Morescoreview} for a detailed up-to-date status of the CC method and a thorough discussion of the systematics uncertainties~\cite{Morescosys} and how to accurately estimate them). 

While the CC method has been known for two decades~\cite{JimenezLoeb}, it has relied so far on spectroscopy, which is expensive in terms of observing time. 
The samples used are thus inevitably small. In the case of CC, statistics is very important as purely passively evolving galaxies (the best CC) are very rare and large samples are needed to truly populate the red envelope, the oldest edge in the age-redshift plane. One alternative which we explore here is to employ high-quality photometry, both in terms of requiring a high signal-to-noise and wide spectral coverage. 

The problem with photometry, especially in the broad-band case, is that it does not provide enough sensitivity to measure the relative ages of galaxies as it integrates over many spectral features that are crucial to obtain the age while marginalising over other parameters, like metallicity and dust. This is illustrated in Fig.~6 and 7 in~\cite{MOPED1}. There, it is shown where most of the signal as a function of wavelength appears for age and metallicity. These same figures give clues on how to design photometric surveys that can measure both features simultaneously. While age sensitivity is mostly weighted towards the ultra-violet and visible parts of the spectrum, the metallicity is spread over the whole visible and infrared wavelength range. 
The question to ask therefore is: to what extent it is possible, with very large wavelength coverage and somehow narrow photometric bands, to obtain robust and precise ages of passive galaxies from photometric surveys?  What would it take to make an $H(z)$ determination from photometric data competitive with other  (emerging) probes?

Here we  present a proof of principle approach to address this challenge.
 This is particularly timely with the almost imminent commissioning of the Euclid satellite and the Vera Rubin Observatory and its LSST survey. 
 In this work, which should be seen as a feasibility study, we set out a method to extract ages from photometry. We do this by using a carefully selected spectroscopic set to train a deep neural network that we then use it on a high-quality photometric data-set to extract the age-redshift relation. 

\section{Methodology}
\label{sec:method}
\subsection{Data}
We take advantage of the rich information available in the COSMOS field, which has been extensively studied both photometrically and spectroscopically.
We define a spectroscopic and a photometric sample as follows. 

{\bf Spectroscopic sample:} The spectroscopic data are crucial to derive accurate age estimates to build our training sample. The Large Early Galaxy Astrophysics Census (LEGA-C \cite{vanderWel2016}) is a public ESO survey of $\sim 3200$ $K_s$-band selected galaxies performed over 130 nights with VIMOS~\cite{Vimos} on the Very Large Telescope in the COSMOS field. One of the greatest strengths of this survey is to provide spectra with both high resolution ($R\sim3500$) and high S/N, thanks to the 20 hr long integration giving an average S/N $\sim 15$ per pixel (0.6 \AA\, in width) for massive galaxies ($M > 10^{11}$ M$_{\odot}$). The second data release (LEGA-C DR2~\cite{Straatman18}) includes 1988 spectra in the redshift range $0.6 < z < 1.0$, and more recently the final third data release has been published, containing 3528 spectra \cite{vanderWel2021}. In particular, we consider the work by Ref. \cite{Borghi1}, which analysed LEGA-C DR2 with a combined analysis of spectra and photometry. In their work, a very pure sample of cosmic chronometers has been extracted by selecting massive and passive galaxies with a combination of photometric and spectroscopic selection criteria. They adopted a combination of NUV-r-J colors to robustly select passive galaxies according to the criterion proposed by Ref.~\cite{Ilbert13}. This selection criterion has been proven to be very effective to select passive galaxies with respect to other ones (e.g. the UVJ criterion), avoiding the contamination by young ($1-100$ Myr) or dust-obscured outliers. To maximize the purity of the sample, galaxies with significant emission lines were further discarded, obtaining a final sample of 140 cosmic chronometers. Ref.~\cite{Borghi1} analysed this sample measuring and modeling a combination of spectroscopic absorption features (also known as Lick indexes) known to correlate with the age, metallic, and $\alpha/Fe$ content, allowing them to derive spectroscopically accurate absolute ages. These absolute ages are mass-weighted stellar ages of the stellar population that makes up the observed galaxies. 
This set of 140 spectra with their associated spectroscopic redshifts and ages constitutes our training (spectroscopic) sample~\cite{Borghi1,Borghi2}. 

{\bf Photometric sample:} For our data sample, we instead considered the pure photometric sample provided by the COSMOS2015 survey~\cite{COSMOS2015}, providing a set of more than half a million galaxies over two square degrees containing accurately calibrated photometry, but also a large number of both broad and narrow bands so as to have enough photometric coverage to constrain the age. The photometric bands that we consider are $B$, $V$, $ip$, $r$, $u$, $zp$, $zpp$, $IA484$, $IA527$, $IA624$, $IA679$, $IA738$, $IA767$, $IB427$, $IB464$, $IB505$, $IB574$, $IB709$, $IB827$, $NB711$, $NB816$, $Ks$, $Y$, $H$, $J$, $Hw$, $Ksw$, and $yHSC$, providing  an observed wavelength coverage from about $350$ nm ($u$) to $4.5\, \mu$ m. This implies that for the rest-frame of the galaxies in the redshift range $0.6 < z < 1.0$ the photometric bands sample the UV age-sensitive features that reside about $200$ nm, all the optical age and metallicity features and then near-IR mass-sensitive lines. The full shape of the spectral energy distribution is well sampled, which contains large sensitivity to the age of the stellar population. It is worth  looking at Fig.~6 and 7 in~\cite{MOPED1} as guidance to where the age and metallicity information comes from as a function of wavelength. To select the most passively evolving objects, we apply to the original COSMOS2015 catalog the same photometric criterion applied to select passive galaxies in \cite{Borghi1}, the NUVRJ selection \cite{Ilbert13}. With this cut-off applied we have a photometric sample of $\sim 19k$ objects that we analyse to obtain the age-redshift relation.

\subsection{Neural Network Fitting Algorithms}

It seems thus natural to proceed in three steps: 1) train an artificial neural network on the training sample to fit the photometric data for the redshifts and mass-weighted stellar ages. 2) Estimate the joint errors on the recovered redshifts and ages and  3) estimate a smooth age-redshfit  relation in the redshift range that is well sampled by the training set, avoiding  extrapolation.  The relation so obtained  yields an $H(z)$  estimate via $H(z)=-(1+z)^{-1}dz/dt$.

The size of this training set is very modest (considering that, as it is standard practice, it will be further split into a training and a validation set), but the advantage is that it samples very well the parameter space that defines the edge of the age-redshift relationship.

This training set (with spectra convolved with the photometric bands of the photometric sample) is  fed to a  multi-layer neural network implemented via the TensorFlow package (see {\tt https://www.tensorflow.org}). The structure of the network consists of three layers: the first one with $32$ neurons and the second with $16$, ending in a linear layer. Activation is set to a {\tt softsign} function. All this is very standard in the artificial neural network fitting algorithm field, so we do not dwell on technical details. Because of the small size of the target (the photometric data)  and training set, a successful model prediction can be achieved with minimal CPU needs. 
We only consider ages for objects that have a recovered redshift between $0.62\le z\le 0.86$,  this being the redshift  range that well is sampled by  the training set. The results  are robust to this specific choice as long as the training set covers a  redshift range  slightly wider than the redshift cuts imposed on the recovered photometric sample. This ensures that  results are not driven by the network  extrapolating beyond the regime where it has been trained.
As it is usual practice, we perform cross-validation (CV), using a randomly-selected 10\% of the training  sample  as validation set, and repeating the procedure 100 times, to assure that there is no over-fitting. We avoid this by stopping when the training and validation sets loss equate each other. This is achieved usually in about $1000$ epochs.

This whole process can be repeated and every run will have different initial random weights for the neural network and therefore, for the same object, different determinations for the age and redshift. Indeed, we repeat the process 25 times and therefore obtain 25 realizations of age and redshift for each element of the sample. The scatter among these 25 runs  yields an error estimate (of the statistical error component) which   we refer to  as realization uncertainty.

An estimate of the statistical uncertainty in the recovered redshifts and ages can be obtained using the moment network technique developed by~\cite{WandeltErr}. In brief, we train a second network, B, that will estimate the posterior variance of the output of the first network A. Then we train Network B  to predict the square of the difference between Network A and the training set as described in~\cite{WandeltErr}.
This produces for each object in the target set  an  estimate of the error-ellipse in the age-redshift plane. We refer to this method to estimate uncertainties as Posterior Moment Networks, PMN.

\subsection{Age-redshift relation and the red envelope}
\label{sec:edge}

\begin{figure}[ht]
\centering
\includegraphics[width=0.49\textwidth]{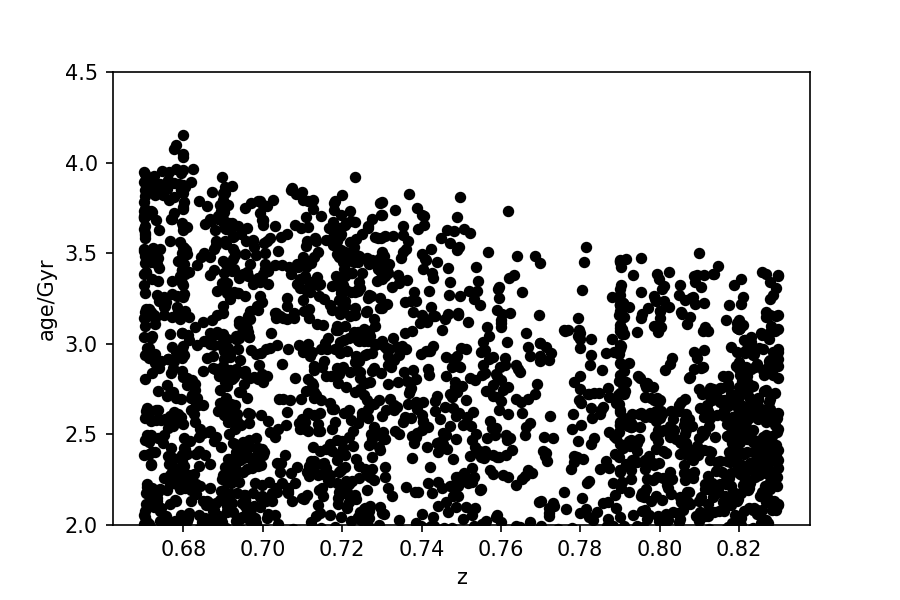}
\includegraphics[width=0.49\textwidth]{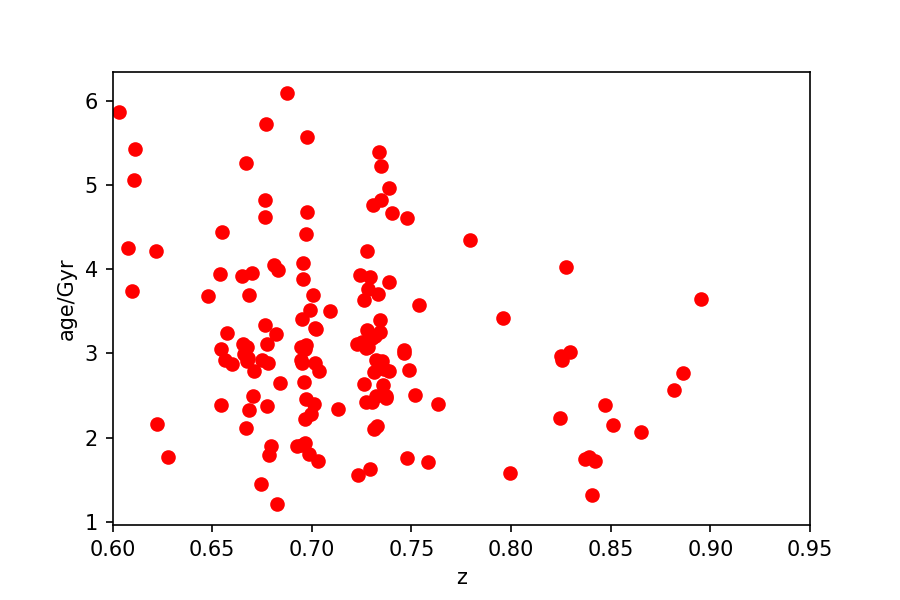}
\caption{Left panel: recovered mass-weighted stellar ages as a function of recovered redshift (black solid dots) for the photometric sample in COSMOS2015, obtained for one of the 25 runs of  the algorithm described in the main text. Right panel: mass-weighted stellar age as a function of redshift (red solid dots) for the spectroscopic sample~\cite{Borghi1} that serves as training set.
}
\label{fig:agevsred}
\end{figure}

Fig.~\ref{fig:agevsred} left panel  shows, for one of the 25 realizations, the age-redshift relation for the photometric sample (solid black dots) obtained from the fit by the neural network using the training set described above. On the right panel, the red points are the spectroscopic age determinations using Lick indexes for the most massive galaxies (with velocity dispersion $\sigma_{*} > 215$ km s$^{-1}$) obtained by~\cite{Borghi1}. 

An age-redshift envelope emerges from the scatter of the points in Fig.~\ref{fig:agevsred}, quite similar to the predictions  for the so called ``red envelope'' in~\cite{Treu+} (their Fig.~5): lower redshift galaxies are older than higher redshift ones.
The shape of the edge of this distribution, the red envelope, encloses information about $H(z)$.

\begin{figure}[ht]
\centering
\includegraphics[width=0.5\textwidth]{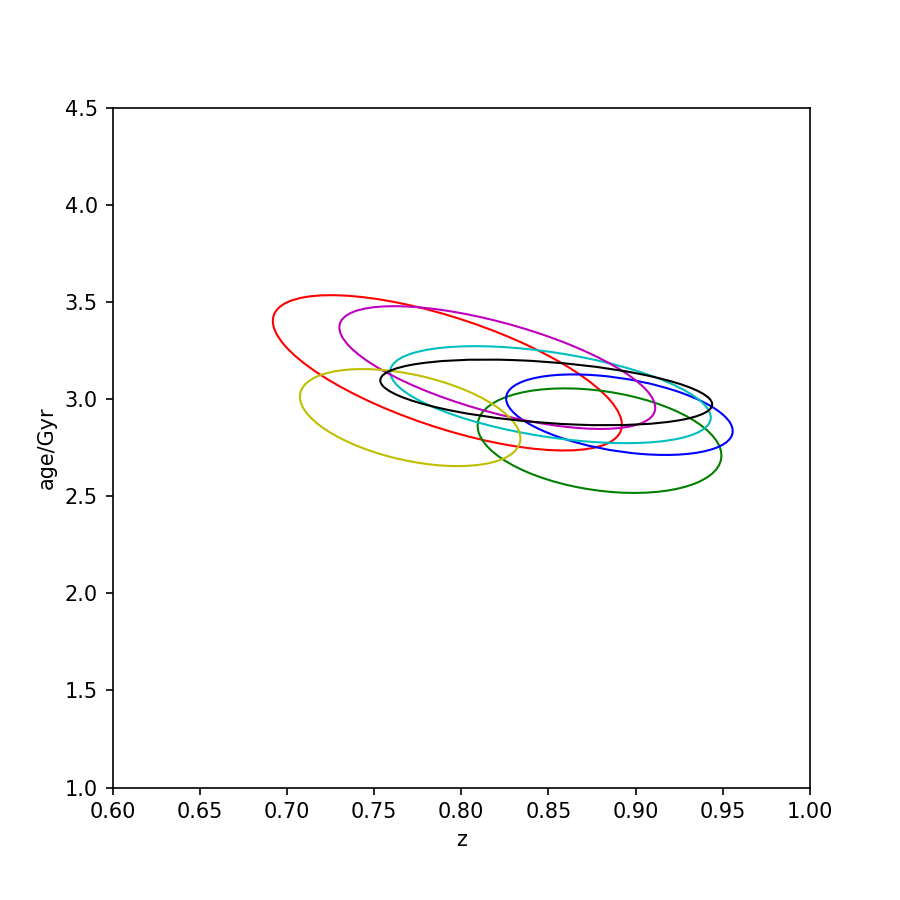}
\includegraphics[width=0.5\textwidth]{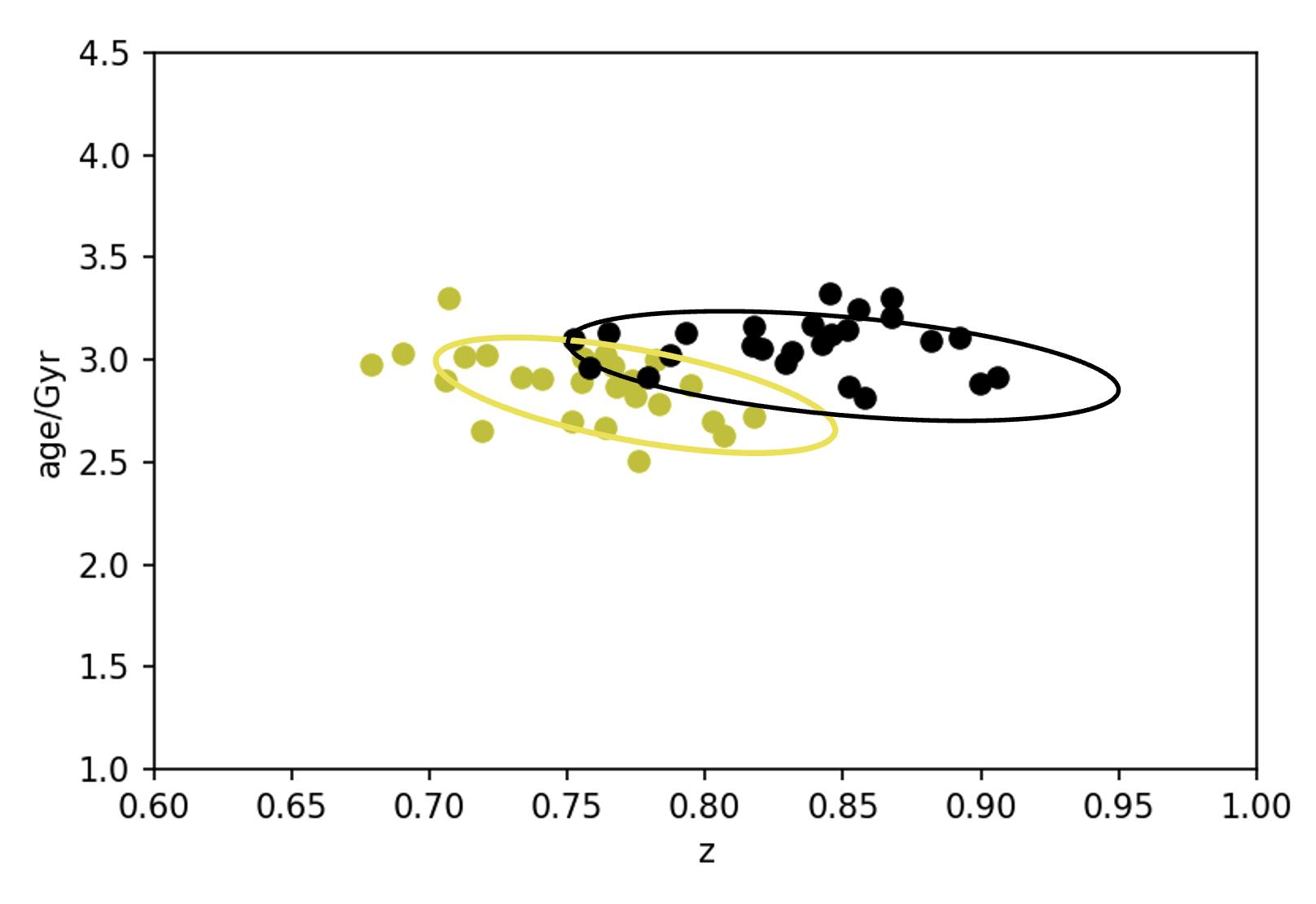}
\caption{Top panel: 68\% join confidence contours for seven representative  individual galaxies to illustrate  the uncertainty in the two parameters we determine: redshift and age. To compute these confidence contours we have used the PMN method described in the text. Bottom panel: same as in the top panel but this time using the realization uncertainty method. We show a couple of examples for clarity and the solid lines  show for reference  the result of the PMN method. Note the good agreement.
}
\label{fig:individual}
\end{figure}

To illustrate the individual joint uncertainty in age and redshift, in Fig.~\ref{fig:individual} we plot the 68\% joint confidence  (PMN method) for seven 
representative galaxies (we limit the number to seven for clarity). There is some degeneracy of trading age for redshift, but it does not spoil the 
red envelope. The confidence contours obtained from the scatter of the 25 realizations  yield an error estimate that is  fully consistent, few typical examples are shown in the right panel of Fig.~\ref{fig:individual}. As the two methods yield very similar estimates of  uncertainties (we estimate $\sim$ 10\% error on the error), below we propagate the realization uncertainty  into the final error budgets.

We now proceed to estimate the red envelope, the edge of the age-redshift relation, by using different quantitative techniques.

\begin{figure}[t!]
\centering
\includegraphics[width=0.7\textwidth]{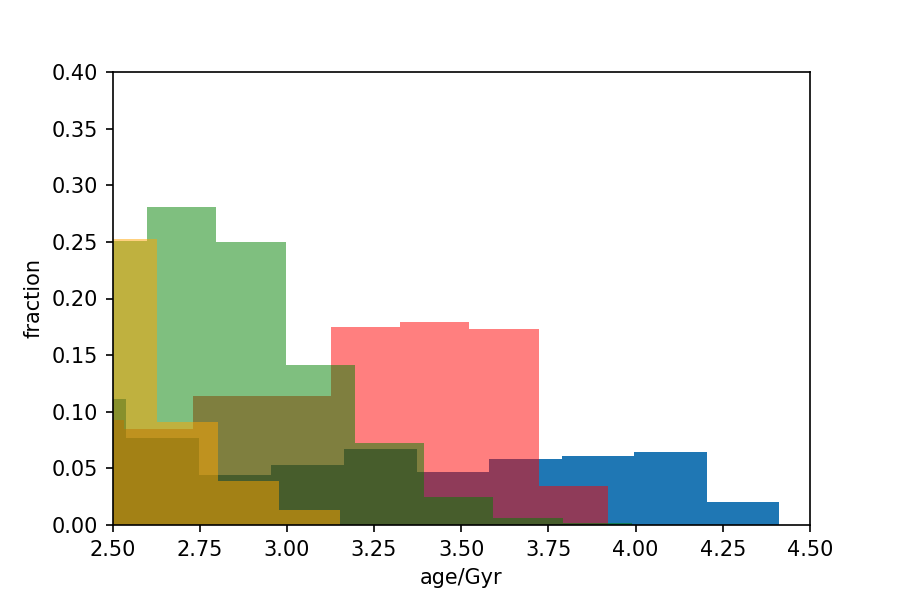}
\caption{Histograms of the age distribution for redshift bins: $0.65 < z < 0.7$ (blue); $0.7 < z < 0.75$ (red); $0.75 < z < 0.80$ (green); $0.80 < z < 0.85$ (yellow).}
\label{fig:histedge}
\end{figure}

\begin{figure}
\centering
\includegraphics[width=0.56\textwidth]{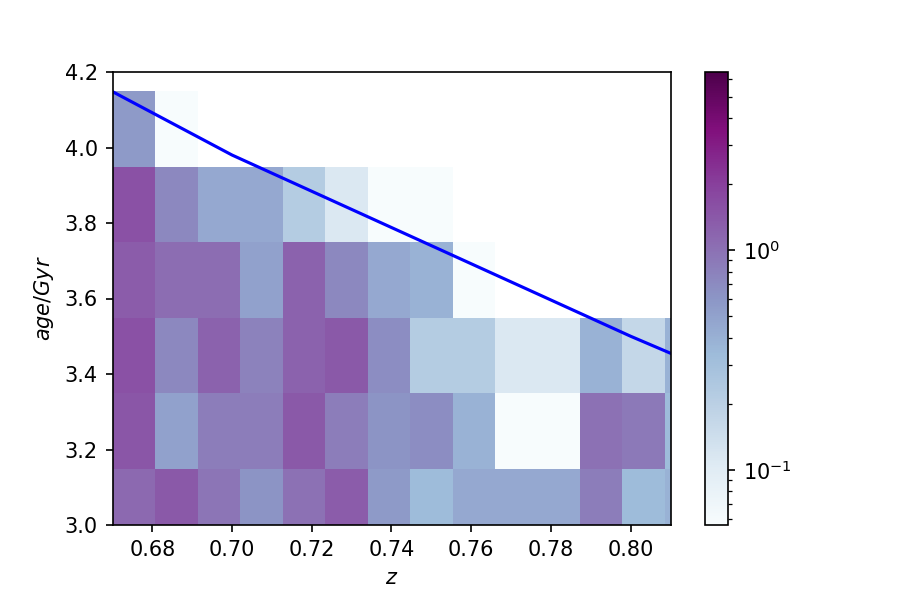}
\includegraphics[width=0.56\textwidth]{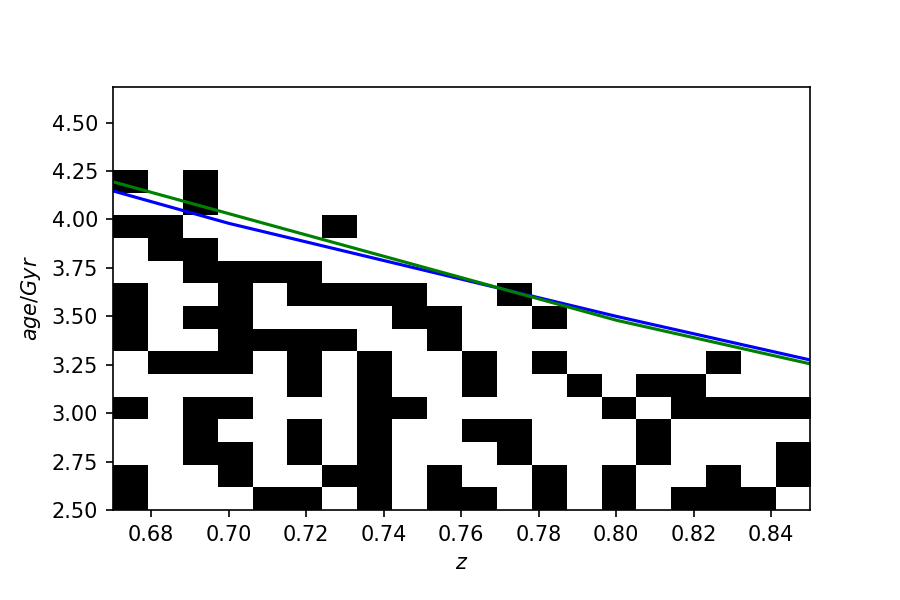}
\includegraphics[width=0.56\textwidth]{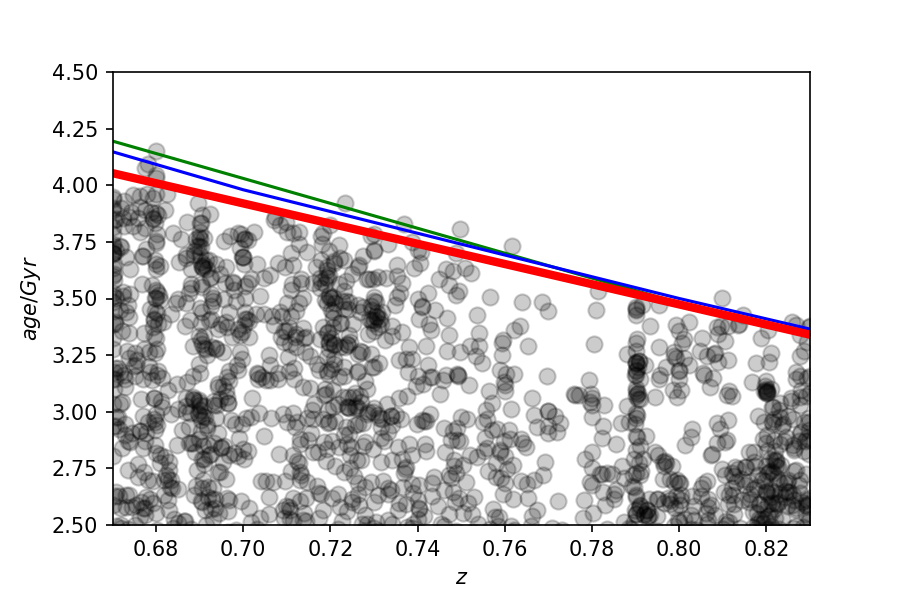}
\caption{The result of three different edge detection algorithms on one of the 25 neural network realizations to find the age-redshift relation of the oldest galaxies. Top panel: Density map of the age-redshift relation and linear best fit (blue line) using the least squared fit of the maximum gradient of the histogram in each of the 13 redshift bins. Middle panel: the Canny algorithm edge detection and best linear fit (green line). The blue line corresponds to  the best fit from the top panel. The internal edges are due to the low number density of points in the sample. Bottom panel: quantile regression. The red line is the linear best fit, the blue line is the best fit from the top panel and the green line is that of the middle panel.}
\label{fig:edge}
\end{figure}

Fig.~\ref{fig:histedge} (which serves only for illustrative purposes) shows the histogram for the recovered age distribution for the photometric sample for the same realization as in Fig.~\ref{fig:edge} in four redshift bins. The age-redshift relation is visible as a shift in the maximum of the histogram.

We use three different algorithms to quantify the edge: 1) a least square-fit to an edge defined as the maximum gradient of an age histogram in suitably chosen redshift bins; 
2) the Canny edge detection algorithm~\cite{canny} and finally 3) quantile regression~\cite{Koenker}. These are initially applied to the age-redshift determinations of Fig.~\ref{fig:agevsred}, without propagating (yet) the errors on these quantities. Results are shown in Fig.~\ref{fig:edge}. 

For the least square fit and the Canny method  we first construct a density map of the age-redshift relation. This is  done by pixelizing the age-redshift plane is $15 \times  13$ pixels (of which only $6\times 13$ are visible in  Fig.~\ref{fig:edge}), and computing a density by simple nearest grid point. This is visible in the top panel of Fig.~\ref{fig:edge}.
For the least square fit method, from  the density map an age histogram is obtained  for  each redshift bin and then the location of the maximum gradient of the histogram is computed, which defines the edge.

 We assume smoothness, as the data do not have enough information to resolve features or curvature of the red envelope  and fit a linear relation of the form:
 \begin{equation}
    {\rm age/Gyr} = a_1 (z-z_p) + a_o\,
\end{equation}
we also adopt $z_p$=0.75, as it is the central redshift of the sample. With this choice the determinations of $a_o$ and $a_1$ are uncorrelated.
 The result of  the least square method for the same realization as in Fig.~\ref{fig:edge} is shown in the top panel of Fig.~\ref{fig:edge} and the blue line shows the best fit. The middle panel shows the result of applying the Canny algorithm to the same density map. In this case, the green line is the best-fitting linear function to the upper edge. There are significant edges inside the age-redshift relation simply because Canny is very efficient at detecting edges and our sample consists of a relatively small number of galaxies (and the density estimation is done crudely with a nearest grid point method). Nevertheless, the upper age is very well defined. Finally, the bottom panel in Fig.~\ref{fig:edge}  shows the result of  quantile regression when including 30\% of the points, the result converges quickly after choosing more than 10\% of the points for the quantile regression. The red line shows the best fit to the edge using quantile regression, while the blue line is the best fit from the top panel and the green line is that of the middle panel. The coefficients for the three methods are: $a_1 = \{-5.31, -4.70, -5.10\} $ and $a_o = \{3.45, 3.78, 3.78\}$ for quantile regression, least square fit and Canny, respectively. This yields an estimate of $\Delta a_1\simeq \pm 0.31$ ($0.25 $ rms) and $\Delta a_o=\pm 0.17$ ($0.16$ rms) to account for the differences among edge detection methodologies. 
 The detailed  behaviour of the $a_o$ scatter  among the three methods, will be studied elsewhere, but we find that the discreteness created by the sharp pixelization in the density estimation, for which the age bin size is  $0.2$ Gy  affects the $a_o$ determination at a comparable level, and indeed it is the same for the Canny and least square fit methods. A different binning choice would yield slightly different edge determination.
 
\begin{figure}
\centering
\includegraphics[width=0.7\textwidth]{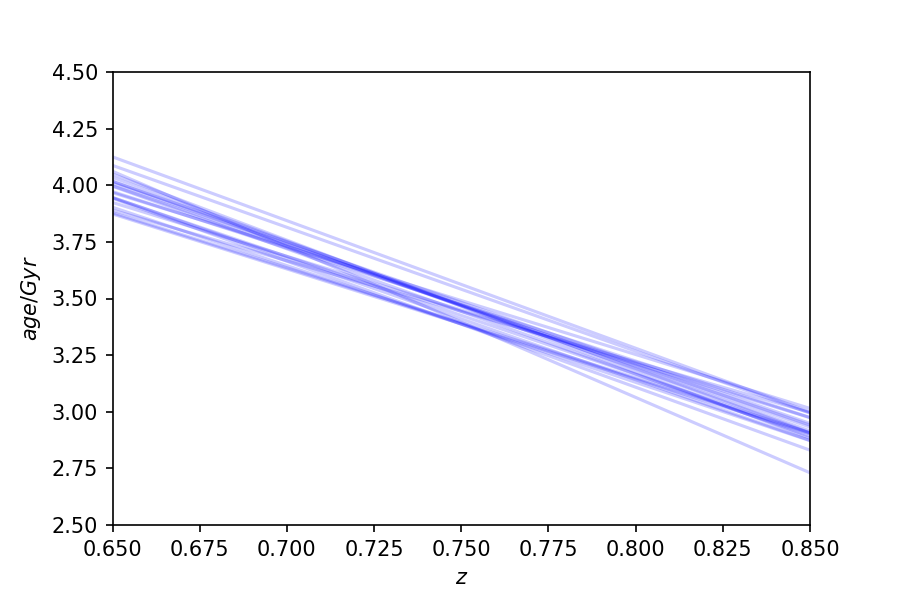}
\caption{Age-redshift relation for the 25 realizations.}
\label{fig:agefit}
\end{figure}

\begin{figure}
\centering
\includegraphics[width=0.7\textwidth]{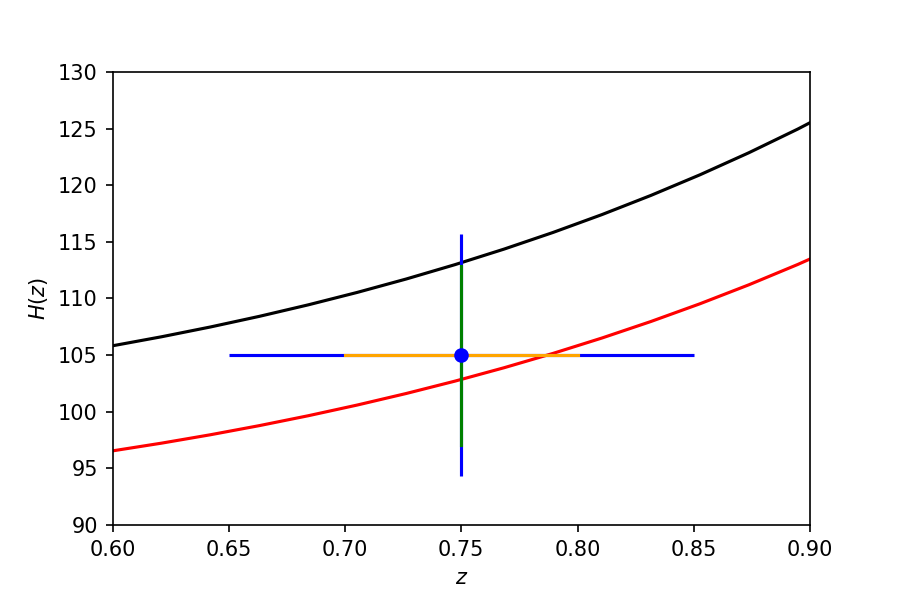}
\caption{$H(z)$ at $z=0.75$ from the photometric CC method (blue dot) and uncertainty region. The blue vertical line is the statistical uncertainty, while the green one is the systematic and statistical uncertainty added in quadrature. The horizontal blue line is the width of all the redshift ranges considered in the sample while the orange one is the typical photometric uncertainty as shown in Fig.~\ref{fig:individual}
The red line is the Planck satellite model~\cite{Planck18} while the black line is a LCDM model with $H_0 = 73$ as in~\cite{Riess} and using the SN Pantheon~\cite{Pantheon+} value for $\Omega_m = 0.334$.}
\label{fig:hz}
\end{figure}

Hereafter, we  adopt the quantile regression as our reference method  as it does not need the intermediate step of  the calculation of a density in the redshift-age plane.  We repeat the quantile regression estimate for the 25 realizations (illustrated by the lines in Fig.~\ref{fig:agefit}) and report the mean and r.m.s. of the coefficients as our red envelope estimate. We find:  $a_1 = -5.31\pm 0.43$ (and $a_o= 3.45\pm 0.05$ although this will be irrelevant for the $H(z)$ inference). With our choice of the pivot point, the coefficients are uncorrelated for all three methods.

\section{Results and Conclusions}
\label{sec:conclu}

From the above,  we can now obtain the Hubble parameter at 
the pivot(effective) redshift ($z_p = 0.75$) provided we can assume that the formation times of all the objects that define the edge of the age-redshift relation coincide, that is, the oldest objects at every redshift constitute a homogeneous population of galaxies that  formed their stellar population at the same time:
\begin{equation}
 H(0.75)=-(1+0.75)^{-1}[a_1]^{-1} 
\end{equation}
and our consensus estimate of the coefficients and their errors is: $a_1 = -5.31 \pm 0.43$.
Therefore
\begin{equation}
H(0.75) =105.0\pm 7.9(stat)\pm 7.3(sys)\,
\end{equation}
where,  following \cite{Morescosys} we  have added 6.6\% of systematic error to the recovered $H(z)$ arising from stellar population model  systematic uncertainties in the training sample.

The result is shown in Fig.~\ref{fig:hz} as a blue dot with the corresponding uncertainty: the green vertical line is the statistical uncertainty, while the blue one is the systematic and statistical uncertainty added in quadrature. The horizontal blue line is the width  of the redshift range considered in the sample, while the orange one is the typical photometric uncertainty in the redshift determination of a single object as shown in Fig.~\ref{fig:individual}.

It is interesting to compare the reconstructed $H(z)$ with the expected one in a $\Lambda$CDM model. The $H(z)$ for the best fit Planck18~\cite{Planck18} $\Lambda$CDM cosmology is shown as red line  in Fig.~\ref{fig:hz}, and the one for $\Omega_m = 0.334$ from the Pantheon plus data~\cite{Pantheon+} and SH0ES ~\cite{Riess}-compatible value of $H_0 = 73$ km s$^{-1}$ Mpc$^{-1}$ is shown as a black line.

While our proof-of-principle cosmology-independent determination of $H(z)$ uncertainty does not allow us to distinguish between late- and early-anchored expansion histories, it is tantalizingly close: with a reduction of a factor two it would be possible to distinguish these two possibilities.

In conclusion, we have presented, for the first time, a  cosmology-independent constraint on $H(z)$ from photometric data. This should be considered as a proof-of-concept or  feasibility demonstration for  extending the Cosmic Chronometers method,  which so far used exclusively spectroscopic data,  to multi-band photometry.  

By exploiting a carefully selected sample with a  large coverage of photometric bands and a small spectroscopic sample that we have used as our training set, we have been able to obtain a clear age-redshift relationship in the redshift range $0.65 < z < 0.85$. This in turn yields a determination of $H(z)$ at an effective redshift of $z=0.75$ with 7\% (10\% when including systematics)  uncertainty.

While the current uncertainty is not yet sufficient to weigh in on the well known ``Hubble tension'',  there are two avenues for progress. First, a better training set should be designed for this specific purpose. Our training set is taken from existing literature and was not developed specifically for this application. The predictions of theoretical stellar population models should be used as a guidance to extend and optimize the training set.
Moreover, half of the error budget of our determination comes from  systematic uncertainties intrinsic to the training set  and specifically the dependence on the choice of stellar population model. This can in principle be reduced significantly by improving the modeling of specific stages of stellar evolution and improving the calibration of  the models to benchmark observations of single stellar populations. 

 The second improvement can be obtained by  using much larger photometric samples such as those  that will be provided by forthcoming surveys (e.g., DESI, Euclid, Rubin, etc.). This larger sample would  increase the redshift coverage and provide more than a determination at a single effective redshift in the near future. It therefore has the potential to have significant impact on the “Hubble tension”  in a way that is independent of currently used approaches and cosmological model assumptions. This will be presented elsewhere.

\begin{acknowledgments}
Funding for the work of RJ was partially provided by
project PGC2018-098866- B-I00 y FEDER “Una manera
de hacer Europa”, and the “Center of Excellence Maria de Maeztu 2020-2023” award to the
ICCUB (CEX2019- 000918-M) funded by MCIN/AEI/10.13039/501100011033. LV acknowledges support by European Union's Horizon 2020 research and innovation program ERC (BePreSySe, grant agreement 725327).
The work of BDW is supported by the Labex ILP (reference ANR-10-LABX-63) part of the Idex SUPER,  received financial state aid managed by the Agence Nationale de la Recherche, as part of the programme Investissements d'avenir under the reference ANR-11-IDEX-0004-02; and by the ANR BIG4 project, grant ANR-16-CE23-0002 of the French Agence Nationale de la Recherche. MM acknowledges the grants ASI n.I/023/12/0 and ASI n.2018-23-HH and support from MIUR, PRIN 2017 (grant 20179ZF5KS).
The Center for Computational Astrophysics is supported by the Simons Foundation. Based on data products from observations made with ESO Telescopes at the La Silla Paranal Observatory under ESO programme ID179.A-2005 and on data products produced by TERAPIX and the Cambridge Astronomy Survey Unit on behalf  of the UltraVISTA consortium.

\end{acknowledgments}

\bibliographystyle{JHEP}

\providecommand{\href}[2]{#2}\begingroup\raggedright\endgroup

\end{document}